# Towards new forms of particle sensing and manipulation and 3D imaging on a smartphone for healthcare applications

*Jinlong Zhu[†\*], Ni Zhao[∥], and Renjie Zhou[‡,§\*]*

[†]Photonic Systems Laboratory, Holonyak Micro and Nanotechnology Laboratory, Department of Electrical and Computer Engineering, University of Illinois at Urbana-Champaign, Urbana, Illinois 61801, USA

[∥]Department of Electronic Engineering, The Chinese University of Hong Kong, Shatin, New Territories, Hong Kong SAR, China

[‡]Department of Biomedical Engineering, The Chinese University of Hong Kong, Shatin, New Territories, Hong Kong SAR, China

[§]Shun Hing Institute of Advanced Engineering, The Chinese University of Hong Kong, Shatin, New Territories, Hong Kong SAR, China

\*Corresponding Authors: zhuwdwz1@illinois.edu; rjzhou@cuhk.edu.hk

## Abstract

Close to half of the world population have smartphones, while a typical flagship smartphone today has been integrated with more than 20 smart components and sensors, making a smartphone a highly integrated platform that can potentially mimic the five senses of humans. Recent advancement in achieving high compactness, high performance computing, high flexibility, and multiplexed functionality in smartphones have enabled them for many cutting-edge healthcare applications, such as single-molecule imaging, medical diagnosis, and biosensing, which were conventionally done with bulky and sophisticated devices. Most of the current healthcare applications are developed based on using the photon-sensitive components, such as CMOS sensors, flash & fill lights, lens modules, and LED lights in the screen, leaving the rest of the smart and high-performance sensors rarely explored. In this Perspective, we review recent progresses in advanced sensors in modern smartphones and discuss how those sensors have great, as yet unmet, promise to offer widespread and easy-to-implement solutions to many emerging healthcare applications, including nanoscale sensing, point-of-care testing, pollution monitoring, etc.

**Introduction**

In this new century, healthcare has revolutionized from depending on expensive and time-consuming laboratory tests to more convenient and immediately available tests to the patients or normal persons. Point-of-care-testing (POCT) is an emerging healthcare field that has been driven by advances in many technology advances, such as lab on chip devices, CMOS sensors, internet of things (IoT), cloud computation, wearable or implantable sensors, etc.[1–3] Other aspects of healthcare include continuous monitoring of environment for pollutant detection, food allergy detection, and daily health monitoring.[4–6] Over the past two decades, significant advances have been made in integrated circuits (IC) and information technologies, which led to the birth and flourishing of smartphones that nowadays have a more profound impact in many ways than any other single devices ever made. Smartphones have gradually become affordable and accessible healthcare solutions at resource limited settings or even during pandemic situations, which has contributed to the popularity of mobile healthcare (mHealth) technology. Smartphones, distinguished from feature phones, have integrated state-of-the-art IC designs, touchscreen operating systems that facilitate wider application software, internet, and multimedia functionality, alongside voice call and text messaging. By the end of 2021, the number of smartphone users worldwide is expected to reach to 3.8 billion users. At the same time, driven by Moore's Law in the semiconductor industry, more and more functional units or sensors have been integrated into smartphones. A variety of sensors, including CMOS cameras, light sensors, proximity sensors, accelerometers, fingerprint sensors, gyroscopes, magnetic sensors, temperature sensors, Hall elements, UV sensors, heart rate sensors, and blood oxygen transducers, have made a smartphone a highly integrated platform that can potentially mimic the five senses of humans (i.e., sight, touch, hearing, smell, and taste), and in many cases with higher sensitivity and less physical limits.[4–8] In recent years, tremendous research efforts have been made based on using smartphone components to tackle many healthcare applications at affordable costs with simple-to-implement solutions. For example, the smartphone imaging capability coupled with a cellular network presents an opportunity for developing portable microscopes for medical diagnosis over the internet.[7,9–17] The screens or flashlights of smartphones can act as programmable light sources for chemical and biological sensing[18–22] and the CMOS sensors can be coupled with spectrometers for sulfur interface study and fruit ripeness estimation.[23–26]

There are also possibilities to achieve many high-level healthcare applications by using the smartphone sensors together with delicate external accessories, e.g., the aplanatic lens module can be used for improving the smartphone imaging performance and deep-learning algorithms can be used to enhance image contrast and resolution.[27–30] As three-dimensional (3D) printers are widely available nowadays, the fabrication of complicated external accessories for smartphones has become easier and cheaper, thus leading more accessories built for diverse applications, such as sensing,[31] point-of-care testing,[32,33] microscopic imaging,[12,34] and environmental monitoring.[35] To the best of our knowledge, most of the demonstrated applications are based on the photonic components in smartphones, such as lens modules, CMOS sensors, LEDs and flashlights, and screens. This may due to the fact that photonic sensors are easier to be integrated with external components and optical systems (e.g., optical microscopes and optical sensors).[12] However, up-to-date there are very few reports on the implementation of other high-performance sensors such as fingerprint sensors and Hall-effect sensors that are based on electron-matter-interactions. Unlike the photon-matter-interactions, the use of electron-matter-interactions will enable new lab-on-smartphone designs to catalyze new forms of healthcare applications.

In this perspective, we review the use of several high-performance sensors in smartphones alongside the combination with external accessories and propose several new avenues to tackle the challenges in m-Health applications. For example, we propose to use ultrasonic fingerprint sensors combined with externally acoustic metasurfaces to manipulate particles in fluidic environments. We will also discuss the development of Hall-effect sensor-based accessories for potential applications in immunoassays. Moreover, the potential use of photonic sensors, including the organic LED (OLED) panels and heartrate sensors, for 3D optical imaging and sensing are proposed and discussed. We envision these frameworks will pave new revenues in the m-Health field.

**Photon-free sensors**

*Ultrasonic fingerprint sensor-based metasurfaces and tweezers*

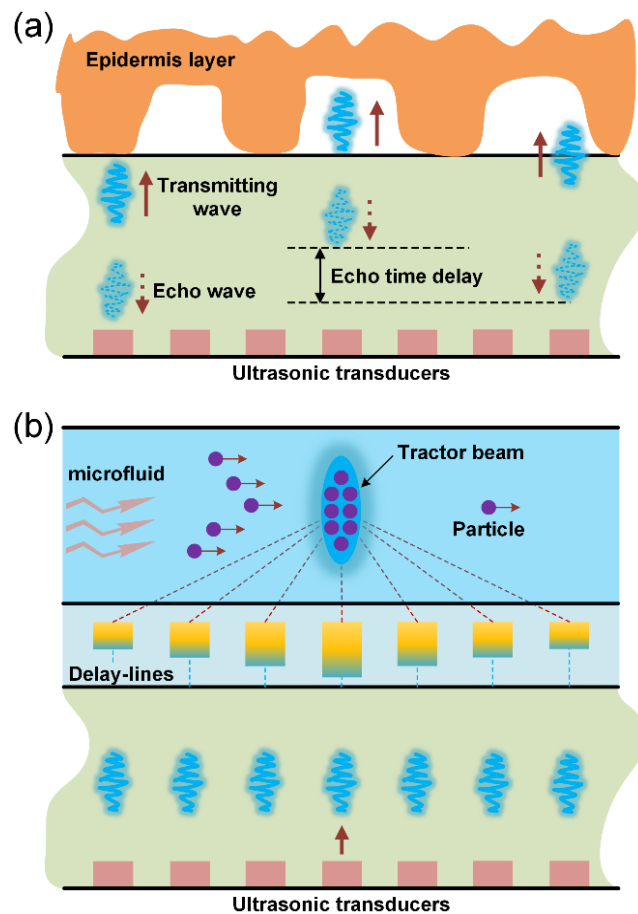

**Figure 1**. Ultrasonic metasurfaces for particle manipulation based on a smartphone fingerprint sensor. (a) Illustration of the principle of the ultrasonic fingerprint sensor based on an ultrasonic transducer array buried underneath the glass panel. (b) Schematic of the ultrasonic tweezer based on a delay-line metasurface for manipulating particles in a microfluidic channel. The delay lines can be either passive or active, depending on the mechanisms of phase modulation in use (for instance, spatially varying solid density or temperature that is tunable via external circuits).

**Operation principle**: The fingerprint sensor is the first used functional component after a smartphone is turned on. Unlike existing photographic or capacitive-based fingerprint sensors that only reproduce low resolution two-dimensional (2D) features of fingerprints, ultrasonic fingerprint sensors make use of very high-frequency ultrasonic sound to scan through the fingerprint and map out the 3D details, including the axial dimensions of valleys and ridges.[36,37] 3D details are much more difficult to forge than 2D, therefore making the ultrasonic sensors much more secure. In an ultrasonic fingerprint sensor (see the schematic in Figure 1a), an ultrasound pulse is first transmitted to the finger that is placed over the screen. Then an echo is bounced back to the sensor, which contains a mechanical stress detection module (transducer array) that can obtain the intensity of the returning pulse at different points. 3D details of fingerprints can be reconstructed from the measured intensity and the echo time delay, that is, the thickness between the epidermis layer (the top yellow objects in Figure 1a) of fingers and the fingerprint sensor glass panel.

**Potential biomedical application**: It can be seen that the core component of the fingerprint sensor is the transducer array buried under the smartphone touchscreen, which generates ultrasonic pulse that can propagate to the glass panel surface to interact with the external environment (Figure 1a). A potential biomedical application of this array is acoustic-wave based particle manipulation. Acoustic-waves, as a form of mechanical waves carrying momentum, can generate radiation forces to levitate particles under certain conditions.[38,39] This can be understood via Newton's second law of motion. Compared with light waves, ultrasound waves can manipulate a wider range of materials, including larger particles and optically opaque substances.[40] To achieve an optical tweezer, a high numerical-aperture objective is used to generate a tightly focused spot with a large intensity gradient for particle trapping. Similarly, a phase array equipped with delay lines can be introduced to generate an acoustic tractor beam, i.e., a focused acoustic beam.[41] Under ideal circumstance, an acoustic focus spot will form with a circularly convergent wavefront. To achieve such a wavefront in front of the smartphone transducer array, an electronic phase modulation method can be used, i.e., customizing the triggering time for each pixel in the transducer array to control the phase delay for each pulse.

Moreover, because each pulse diffracts into the free space, only a small portion of the wave constructively interferes in the focal region, giving rise to a weak focal spot.

To solve those issues, we can alternatively use external metasurfaces (or delay lines) to delay the in-phase pulse arrays as illustrated in Figure 1b. As ultrasound devices usually operate at frequencies ranging from 20 kHz to several GHz, a typical consumer 3D printer can be used to fabricate the external delay lines with sufficient printing resolution. The delay lines can be either active or passive, depending on the method used to control the phase. For example, as the speed of acoustic wave depends on temperature and solid density, we can either use an external temperature-control circuit or a passive solid with a predesigned density via effective medium theory to tune the phase at the output channel of each pulse. The governing equation for achieving a steady focal spot from a circularly convergent wavefront is,[42]

$$\varphi(x,y) = \frac{2\pi}{\lambda}\left(f - \sqrt{x^2 + y^2 + f}\right) \quad (1)$$

where $\varphi$ is the phase delay of the ultrasonic wave at each delay line, $(x, y)$ denotes the position of the delay line, $\lambda$ is the wavelength of the ultrasonic wave, and $f$ is the designed focal length. A trapping potential then can be generated by the phase delay lines to levitate particles, e.g., $SiO_2$ particles and viruses either in fluidic environment or in the air (Figure 1b). Franklin *et al* demonstrated the trapping and manipulation of polystyrene micro-particles in the Rayleigh regime (radius from $\lambda/14$ to $\lambda/7$) using a 3D-printed Fresnel lens with a peak trap stiffness of 0.43 mN/m at the focus (Form 2, Formlabs, MA, USA) and a two-dimensions piezoelectric transducer (PZT-5H, Beijing Ultrasonic, P.R. China).[40] Although the transducer they used is not the one in smartphone, their preliminary investigation has provided a solid foundation that acoustic trapping can be potentially achieved in mobile devices as the piezoelectric transducer and smartphone fingerprint sensor have the same working principle in terms of manipulating acoustic waves.

The realization of particle manipulation can be link to a range of potential applications, such as point-of-care testing,[43,44] water quality monitoring,[45] and micro-robotics.[46] The stiffness and efficiency of trapping will be affected by the quality of 3D printed phase delay lines. If, for instance, the bottom surface of the printed device is tilted or uneven, the generated focal spot will be distorted, thus leading to a weak trapping potential. The transducer array can be used as a surface quality monitor for the printed device. For a perfect bottom surface that well contacts the glass panel, the reflectance of the ultrasonic pulses captured by the transducer array should be uniform (approaching 100%). If there is an air gap, resulting from the tilted or rugged bottom surface of the printed device, there will be an impedance mismatch between air and the glass panel that can potentially result in a strong reflection of nearly 100%.[36] By analyzing the reflectance distribution at the interface, the surface quality of the printed phase delay device can be quantitatively characterized.

*Hall-effect sensor for immunoassays*

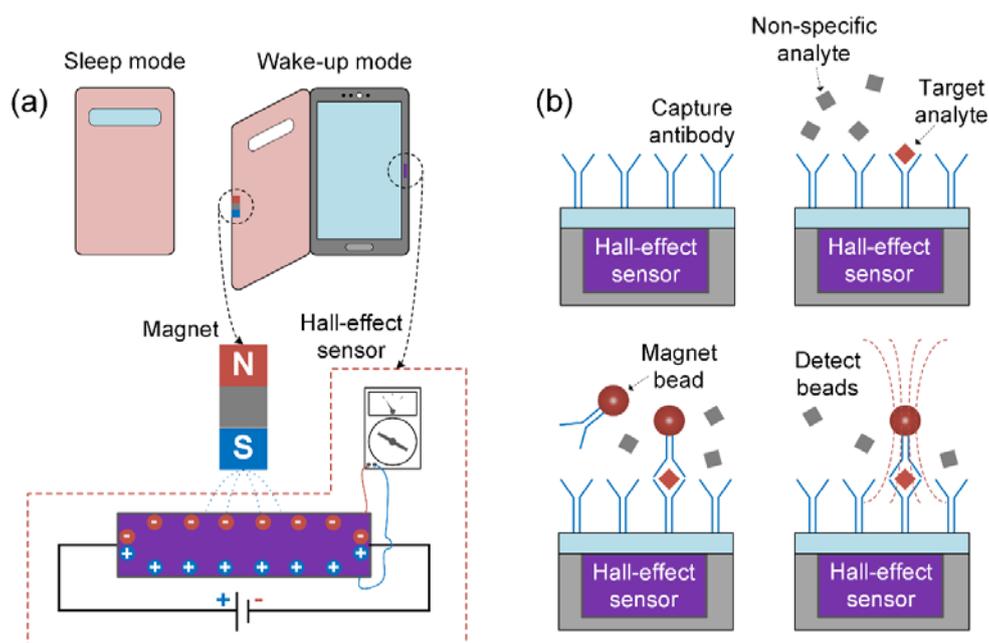

**Figure 2**. Magnetic immune-assay based on smartphone Hall-effect sensors and magnetic beads as labels. (a) Illustration of the principle of Hall-effect sensor and its application in smartphone wake-up services. (b) Illustration of the principle of magnetic immune-assay: the smartphone Hall-effect sensor surface is functionalized with a capture antibody (1) and exposed to the sample. Antibody labelled with magnet beads is subsequently introduced into the chamber (3). Magnetic field generated by the bounded beads perturbates the Hall-effect sensor and thus results in a signal of bias voltage.

**Operation principle**: Hall-effect sensors have been used in smartphones to lock or disable the screen whenever a flip cover equipped with magnets touches the screen. Hall effect (named after the American physicist Edwin Herbert Hall) states that a voltage can be measured at right angles to the current path when a conductor or semiconductor with a current flowing in one direction is introduced perpendicular to a magnetic field.[51] Figure 2a shows a smartphone Hall-effect sensor alongside a schematic illustration of the working principle—charges flow in a thin metal strip (a key component in the smartphone Hall-effect sensor) accumulate on the top and bottom surfaces due to the Lorentz force on charges, induced by the externally magnetic field of a magnet mounted in the smartphone cover. The strength of external magnetic field is proportional to the surface voltage, which can be measured by a voltmeter (see the bottom inset in Figure 2a).

**Potential biomedical applications**: Hall sensors can be utilized for sensing external change of magnetic fields via voltage measurement, which paves the route to achieving biosensing or nanoscale sensing, other than using light, plasmonics, and thermal-effect-based sensing platforms.[47–49] Moreover, Hall-effect sensors can be scaled to several micrometers in size, which lays the foundation for portable and integrated biosensing.[47] Consider an immunoassay example as shown in Figure 2b, where magnetic beads are used as labels on a Hall sensor. Magnetic beads are suitable as labels since their signals are stable in biological systems and buffer solutions, thus allowing for robust detection and long shelf life.[47] First, the surface of the Hall sensor is functionalized with antibodies that specifically bind to target analytes being detected. Second, an aqueous sample containing the target analytes is introduced and bind to the surface via the capture antibodies. Subsequently or concurrently, magnetic bead labels are introduced and bind to the target analytes via conjugated detection antibodies specific to another section of the target analytes. Any non-specifically bound beads may be washed away in fluidic systems. The remaining bound beads can then be detected because their intrinsic magnetic fields can change the current via accumulating charges on vertical surfaces of the conductor in the Hall-effect sensor.[47,50] Based on such concept, a magnetics-based immunoassay prototype has been demonstrated using commercial or customized complementary metal-oxide-semiconductor (CMOS) Hall-effect sensors.[48,50] The quantification of 1% surface coverage of 2.8 μm beads has been achieved within 30 seconds.[47] These preliminary investigation has demonstrated that CMOS Hall-effect sensors, which are also widely used in smartphones, indeed have the potential to sense micrometer-scale magnetic beads for high sensitivity biosensing. The fluidic chamber containing magnetic beads and analytes can be fabricated using commercial 3D printers. To fully control the Hall-effect sensor for quantitative analyte sensing, one may develop a software application or a third-party software. The response sensitivity of the sensor determines the minimal detectable dosage of magnetic variation (i.e., number of magnetic beads). This sensitivity may need to be precisely calibrated before actual sensing—for example, one can use calibrated magnetic beads with various magnetic field strengths to test the threshold of Hall-effect sensors.

**Photonic sensors**

*OLED panel for 3D label-free imaging of cells*

**Operation principles**: Feature phones equipped with liquid crystal displays (LCDs) dominated the market until 2010. A spatial light modulator (SLM) consist of a 2D array of pixels are made from liquid crystal molecules. By reverse engineering of LCDs in feature phones, researchers have designed cost-effective SLMs that can be used in optical imaging and biomedical optics applications.[51,52] Nowadays, organic light-emitting diode (OLED) panels have been dominating the smartphone display market. While LCD panels offer boast rich color, detail, and contrast, OLED display is the first display technology that allows self-lighting pixels to be switched on and off individually. OLED panels have been recently implemented as illumination sources for optical biosensors[53–55], but this is far from taking their full advantage.

A typical smartphone OLED panel structure is illustrated in Figure 3a, where one pixel represents the smallest repeating pattern that contains a minimal number of RGB elements (see the inset in Figure 3a). Each pixel can be made as small as 100 μm × 100 μm, and more than two million pixels can be integrated on a smartphone OLED panel.

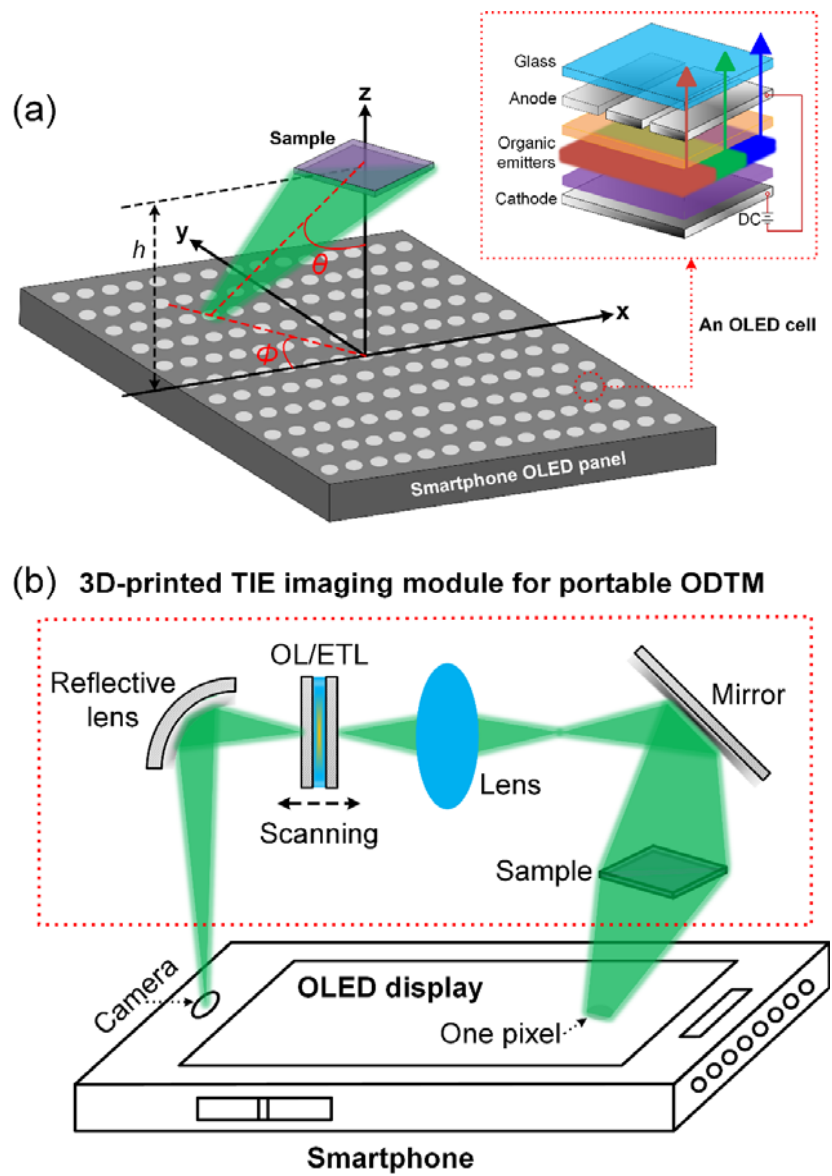

**Figure 3**. 3D-printed TIE imaging module for portable ODTM. (a) Schematic of the OLED panel-based illumination system. The smartphone OLED panel is placed under the sample at a given distance $h$. The inset denotes the schematic structure of an OLED cell. (b) 3D-printed TIE imaging module that consists of a plastic shell (represented by the dotted red box), a reflective mirror, an imaging lens, an OL/ETL module for fast axial scanning, and a reflective lens for guiding the imaging field to the built-in front-facing camera.

**Potential biomedical applications**: Now we consider such panels utilized as an illumination source in an optical imaging system and the on and off sequence of each pixel is programmable. By placing the specimen in proximity to the panel, an ultrawide-angle illumination cone and angle-resolved illumination mode can be achieved.[56] In order to increase light brightness for illumination, one may use a group of pixels rather than one single pixel. This makes it possible to realize optical diffraction tomography (ODT) for 3D live cell imaging by mapping the 3D refractive index distributions.[57–59] However, one may be concerned about the temporal coherence of the OLED source as it may affect the phase measurement accuracy in ODT. Conventionally, narrow-band light sources are preferred as the phase is well defined, making the reconstruction of the refractive index maps of specimens more accurate. Recent studies have shown that phase retrieval through the transport-of-intensity equation (TIE) method under a partially coherent illumination can also be used to reliably infer the refractive index distribution of specimens.[60–62] Therefore, we postulate that an OLED panel can be potentially utilized as a scanning light source for ODT by combining with the TIE-based phase retrieval algorithm. Such implementation may have advantages over conventional approaches as the speckle noise can be reduced with partially coherent sources and environmental vibrations and mechanical error during data acquisition process can be also minimized.[63,64] Moreover, the high integration level of pixels in an OLED panel makes it possible to design a highly miniaturized and cost-effective instrument. As each pixel contains red, green, and blue color channels, one may even reconstruct the refractive index distribution of a specimen at three different wavelengths simultaneously by mounting three bandpass filters on three independent channels respectively in the imaging space. The filters also can improve the coherence of the detected signals without introducing significant speckle noises.

In a simplified schematic design shown in Figure 3b, the smartphone front-facing camera is combined with an OLED panel to enable us to build a low-cost miniaturized optical diffraction tomography microscope (ODTM). The scattering field coming from the sample under angle-resolved illumination is projected by a mirror to the back-focal plane (BFP) of a lens. An electronically tunable lens (ETL) combined with a plane concave offset lens (OL) is mounted right at the BFP of the former lens. The OL/ETL module enables a rapid electronical control of the focal plan so that we can capture three or more through-focus frames. A reflective lens is introduced to guide the imaging field into the front-facing camera of the smartphone. Phase can be reconstructed from the through-focus frames by a TIE solver.[63,65,66] With the recovered amplitude and phase maps corresponding to different illumination angles, the 3D refractive index distribution of the specimen can be reconstructed using an inverse scattering model.[59,60] Note that all the passive components in the ODTM can be packaged in a 3D printed cage. 3D printed scaffolds of passive optical components, including lenses and mirrors, have been implemented in various types of mobile optical microscopes, such as brightfield, darkfield, immersion, fluorescence, quantitative phase, and even lensfree microscopes.[8,10–12,15,32] Therefore, it is practical to achieve the proposed mobile ODTM. We envision ODTM, as a low-cost and portable 3D label-free imaging modality, will find applications in bioimaging and point-of-care testing at resource limited settings.

*Heart-rate sensor for micro- and nanoscale sensing*

**Operation principle**: Heart-rate sensors are currently equipped in most flagship smartphones.[67–69] A mainstream heart-rate sensor measures the heart rate in beats per unit time interval using an LED source and a photodetector; see the picture of a standard heart-rate sensor on the left of Figure 4a, where three photodetectors at three different positions are used to guarantee accurate heart rate measurement under motion state. When the heart is beating, blood pressure pulses are generated and propagate in the blood vessel, leading to an increased local pressure. This can modify the geometry and the property of the medium in the blood vessel, which will result in an increase of light absorption, thus an attenuation of light transmission (see the principle of heart-rate sensor illustrated in Figure 4b). By continuously monitoring the dynamics of the blood vessel, periodic reflectance or transmittance spectrum can be obtained. By counting the number of pulses in a unit time interval, the heart rate can be obtained (see a representative time-resolved signal in Figure 4a).

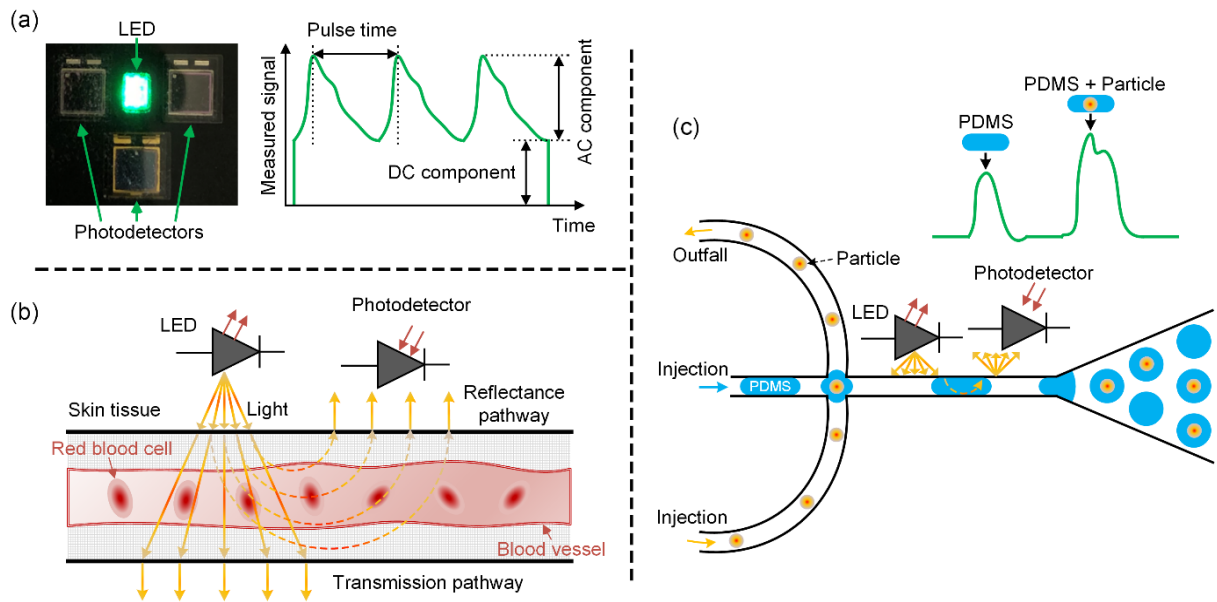

**Figure 4**. Microfluidic device integrated with a smartphone heart-rate sensor for particle isolation and detection. (a) A representative heart-rate sensor consisting of a LED light source and photodetectors (left) and a simplified representation of the measured signal on the photodetectors (right). (b) Light reflectance mode versus transmission mode in heart-rate sensors. (c) Schematic showing the application of smartphone heart-rate sensor in a microfluidic device for monitoring flow dynamics. The module can be used to count and distinguish the PDMS droplets with and without particles (i.e., viruses, cells, metallic particles, etc.) by recording the differential reflectance peaks or valleys in the measured signal (see the inset at the top right corner).

**Potential biomedical applications**: The fundamental principle of heart-rate sensor is the nonzero reflectance induced by the nonzero refractive index difference on the fluid-cell interface. This indicates that heart-rate sensors can be applied to the monitoring of other dynamic processes that can result in optical signal variations. Now consider a microfluidic system in which microdroplets are injected from left to right (see Figure 4c). The microdroplets can be made of polydimethylsiloxane (PDMS), which is widely used in cell circadian rhythms studies and microfluidic chips.[70–73] The goal of the microfluidic system is to isolate individual particles (e.g., viruses, cells, metallic nanoparticles, etc.) injected from the bent pipe using

PDMS. In the ideal case, each particle should be isolated via a PDMS shell. However, due to the error in synchronizing the PDMS droplets and particles at the intersection point of the straight and bent pipes, there is a chance that a vacant PDMS droplet will be created and injected into the output channel (the left opening where all droplets leave the system). The goal is to detect and classify PDMS droplets with and without embedded particles.

When the PDMS droplets are illuminated by an LED source, the refractive index difference between droplets and ambient liquid will result in a nonuniform back reflection whose strength depends on the incident angle of the light ray on the interface. For instance, the reflection at the water-PDMS interface at 632 nm light under normal incidence is 0.085%. Therefore, for a high-performance heart-rate sensor in which the power of the LED emitter is more than 4 W, the reflection strength can reach more than 3.4 mW, which is sufficient to be detected by an LED detector whose detection sensitivity can be as low as microwatt level. The angle-resolved reflectance on a water-air interface is then recorded by an LED light detector. As the liquid flows in the tube, periodic reflectance as a function of time can be obtained, similar to that in the monitoring of blood vessel dynamics using the heart-rate sensor. If a PDMS droplet contains a particle, there will be an increase or decrease of light backscattering, thus leading to an increased or decreased in the detected time-resolved signal (see the schematic in the top right corner of Figure 4c). The difference in the peak signals will enable us to determine whether a PDMS droplet contain particles or not. Counting the number of a specific type of peaks in the spectrum in a unit time interval will give the number of droplets generated in the system. This design shows that a heart-rate sensor can be utilized as a feedback control of a dynamic microfluidic system for micro- and nanoscale sensing for healthcare applications.

**Conclusion**

New breakthroughs in semiconductor technologies are actively driving the miniaturization of diverse electrical, acoustic, and optical sensors that can be seamlessly integrated into smartphones. A modern smartphone not only acts like a tiny computer, but also on the way to fully mimic the five senses of humans (i.e., sight, touch, hearing, smell, and taste) in perceiving the external environment via diverse advanced sensors. It is the high integration, portability, and low cost that makes smartphones a more and more attractive POCT platforms. In this perspective, we have reviewed, proposed, and discussed new implementation of recently available non-optical sensors (ultrasonic fingerprint sensors and Hall-effect sensors) and unconventional optical sensors (OLED panel and heart-rate sensors) for biomedical applications including particle manipulation, immunoassays, 3D label-free imaging of cells, and biosensing. Our goal is to demonstrate that there is an endless potential of the integrated sensors in smartphones that has not been fully explored in past, which will significantly extend the territory of "lab on a smartphone." We believe the emerging 3D printing techniques, new functional materials, alongside the never-ending pursue of advanced smartphone sensors that can mimic our five senses or even beyond may catalyze more and more unprecedented applications.

**Acknowledgements**
This work was funded by Cisco Systems Inc. (Gift Awards CG 1141107 and CG 1377144); University of Illinois at Urbana-Champaign College of Engineering Strategic Research Initiative; Zhejiang University – University of Illinois at Urbana-Champaign (ZJUI) Institute Research Program; Hong Kong Innovation and Technology Fund (ITS/394/17, ITS/098/18FP); Shun Hing Institute of Advanced Engineering (BME-p3-18); Croucher Innovation Awards 2019.

**Author Contributions**
J. Zhu and R. Zhou conceived the concept and wrote the manuscript. Ni Zhao supervised the project.

**Corresponding authors**
Correspondence to Jinlong Zhu or Renjie Zhou.

**Notes**
The authors declare no competing interests.

**Abbreviations**
CMOS, Complementary Metal-Oxide-Semiconductor Transistor; LED, light-emitting diode; UV, ultraviolet; SLM, spatial light modulator; LCD, liquid crystal display; OLED, organic light emitting diode; ODTM, optical diffraction tomography microscope; BFP, back-focal plane; ETL, electronically tunable lens; OL, offset lens; TIE, transport-of-intensity equation; PDMS, Polydimethylsiloxane.